\documentstyle[12pt,fleqn]{article}    
\begin{document}           

\begin{center}
\hfill Warsaw University Report IFD/3/92 (1992) 

\hfill High Energy Nuclear Collisions Group\\

\end{center}

\vspace{3cm}

\begin{center}
{  \Large \bf Search for  Short Lived Particles in High Multiplicity
Environment }\\
\end{center}

\vspace{0.5cm}
\begin{center}

{\bf Marek Ga\'zdzicki\footnote{present address: Institut f\"ur
Kernphysik, University of Frankfurt, August Euler Str. 6,
D--60486 Frankfurt, Germany}
and Waldemar Retyk \\ }

Institute of Experimental Physics, Warsaw University,\\
ul. Ho\.za 69, PL--00-681 Warsaw, Poland \\ 
\end{center}
\begin{center}
and \\
{\bf Jan Pluta
 \\ }

Institute of Physics, Warsaw University of Technology,\\
ul. Koszykowa 75, PL--00-662 Warsaw, Poland \\ 
\end{center}

\vspace{2cm}
{\bf Abstract.} 
A method of statistical selection of short lived particles in
high multiplicity nucleus--nucleus collisions is discussed.

\newpage

\section{Introduction}
High energy nuclear collisions  will produce many
thousands  charged particles in the central rapidity interval 
usually covered
by tracking detectors. Among the measured charged hadrons are those
which are:
\begin{itemize}
\item
directly produced in the process of strong interaction between
two nuclei, 
\item 
products of decay of resonances during
the hadronization process and
\item
products of weak decays of heavy hadrons.
\end{itemize}

The characteristic distance between emission points of particles 
originating from the first
two  processes is of the order 1 to 100 fm; this distance can be
measured only in the statistical way using particle interferometry.
However, the characteristic 
distance between weak decay point and the collision region
is often well measurable using recent experimental technics.

In this note we discuss a possible method of statistical
selection of weakly decaying particles in the experimental
case when the characteristic decay distance is of the same
order as experimental resolution of its measurement. In this case
the main problem is to make
efficient rejection of background due to the particles
originating directly from the interaction.

\section{General Description}

We consider here two body decay of a given particle and assume
that momenta and charges of both decay products are measured
by the experimental set up. Typical examples are neutral
strange or charm particle decays into charged particles 
(V topology) which
can be easily measured.

The basic points of the method are the following:
\begin{itemize}
\item
for each combination of two tracks coming from the same interaction,
which are potential candidates for products of the decay
the hypothetical decay point is fitted and 
the kinematical fit is done.
\item 
 values of  cut--variables are calculated for each pair
and the probability density function (pdf) for accepted pairs
from the same interaction in the cut--space (see below) is obtained 
('raw signal' pdf), 
\item 
the 'raw background' pdf in the cut--space is calculated using
combinations of tracks from different events,
\item
the acceptance region in the cut--space is selected using 'raw background'
pdf, expected 'signal' and 'background' multiplicities and number
of collisions,
\item 
the 'signal' in the acceptance region is  calculated as a
difference between 'raw signal' and 'raw background' pdfs
multiplied by a 'raw signal' average multiplicity.
\end{itemize}

The important element of the method is an introduction of the cut--space,
using which transparent and well defined method for the selection of
the best acceptance region (cuts) for the given experimental
case is developed. 
The procedure should be used separately for different $y-p_T$
regions. The weighting factors, due to applied cuts in the cut--space
necessary to obtain final results are trivial.

\section{Cut--variables }

In this Section we describe a set of five cut--variables which are later
used to select the best acceptance region for  separation  of 'signal' from
the 'background'.
The cut--variables are defined in a specific way, which grantees
several important features of these variables calculated for
the 'signal' (real V decays):
\begin{itemize}
\item
the variables are independent from each other,
\item
the variables are independent of kinematical properties of the decaying
V particle (eg. y or p$_T$),
\item
the  'signal' pdf in cut--variables is independent of an
experimental resolution,
\item
the 'signal' pdf is uniform; the variables range between 0 and 1.
\item
the cuts in the cut--variables do not affect acceptance in the physically
important variables eg. rapidity or transverse momentum.
\end{itemize}

In the following a specific set of 5 cut--variables is described.

\begin{enumerate}
\item
The cumulative $\chi^2$ of the decay vertex fit.\\
The cumulative $\chi^2$ variable of the vertex fit is defined as:
\begin{equation}
C_{V} = \int^{\chi^2}_0 f(u,\nu)du,
\end{equation}
where $f(u,\nu)$ is chi--square probability density function for
$\nu$ degrees of freedom. The $\chi^2$ value for each decay is
calculated by decay vertex fit procedure.
\item
The cumulative $\chi^2$ of the kinematical fit.\\
The cumulative $\chi^2$ variable is defined as:
\begin{equation}
C_{K} = \int^{\chi^2}_0 f(u,\nu)du,
\end{equation}
where $f(u,\nu)$ is chi--square probability density function for
$\nu$ degrees of freedom. The $\chi^2$ value for each decay is
calculated using
standard method of Least Squares estimation with  constrains
(kinematical fit).
In a case of measured momenta vectors of both decay products and
polar, $\theta_V$, and azimuthal, $\phi_V$, angles of the decay point
(defined in the spherical system with the origin in the interaction point)
the number of degrees of freedom is $\nu$ = 3. In the limit when there
is no information on $\theta_V$ and $\phi_V$ $\nu$ = 1.
\item
The polar angle of the decay product.\\
This cut--variable is defined as:
\begin{equation}
C_{\theta^*} = 0.5 (cos\theta^* + 1),
\end{equation}
where $\theta^*$ is an angle between momentum vector of positive (or
negative)  decay product and  V momentum vector calculated in the V
center of mass system.
 The values obtained from the kinematical fit are used for the
calculation.
\item
The azimuthal angle of the decay product.\\
This cut--variable is defined as:
\begin{equation}
C_\phi = \frac {\phi} {2 \pi}
\end{equation}
where $\phi$ is an angle between a projection of the beam direction
and the projection of the momentum vector of the positive (or
negative) decay product onto plane perpendicular to the V direction.
The values obtained from the kinematical fit are used for
calculations.
\item
The cumulative life time.\\
The cumulative life time cut--variable is defined as:
\begin{equation}
C_\tau = \int^{R(\tau)}_0 P_{p_V}(R') dR',
\end{equation}
where $R$ is a distance between measured
V decay point and the interaction point,
$\tau$ is a V life time in the V 
center of mass system and $P_{p_V}(R)$ is a
probability density  function of decay at the distance $R$ calculated
for a V with momentum $p_V$. In the case of an ideal measurement
of the decay distance $R$ one gets:
\begin{equation}
P_{p_V}(R) = P_{p_V}^I(R) = Ae^{- \frac {\tau(R)} {\tau_0}} =
Ae^{- \frac {R} {R_0}},
\end{equation}
where $R_0 = p_V \tau_0 / m_V$ and $A$ is normalization factor.
If the decay distance measurement is not an ideal one we get:
\begin{equation}
P_{p_V}(R) = \int^{\infty}_0 P^I_{p_V}(R') \rho(R,R')dR',
\end{equation}
where $\rho(R,R')$ is a probability density that decay at a distance
$R'$ is measured as a decay at the distance $R$. The values obtained from 
kinematical fit are used for the calculations.
\end{enumerate}

The choice of the given set of cut--variables is not unique, especially
various definitions of angular variables are possible.
The final selection depends on the experimental situation.
The variables should be defined in order to maximize nonuniformity
of the 'raw background' pdf.

\section{Cut-space and Acceptance Region}

We define a 5--D cut--space as a set of points:
\begin{equation}
{\bf C}  = (C_{V}, C_{K}, C_{\theta^*}, C_{\phi}, C_{\tau}),
\end{equation}
From the definitions of $C_i$ given in previous Section follows
that:
\begin{equation}
V = \int \int \int \int dC_{V}dC_{K}dC_{\theta^*}dC_{\phi}dC_{\tau} =
    \int dV = 1.
\end{equation}
For the 'signal' pdf we obtain:
\begin{equation}
\rho_s({\bf C}) = const({\bf C}) = 1
\end{equation}
where $\rho_s({\bf C})$ is a 'signal' pdf at point
{\bf C}.

The 'background' pdf, $\rho_b({\bf C})$,
is obviously nonuniform and therefore the ratio of the 'signal'
to the 'background' depends on the point {\bf C} or on the
subspace selected in the cut--space.

The average number of 'signal' events 
(averaged over all collisions) in the subspace of the volume
$V^{acc}$ around any point {\bf C} is given by:
\begin{equation}
<n_s^{acc}> = V^{acc} <n_s>,
\end{equation}
where $<n_s>$ is an average number of all 'signal' events.
eg. all charged decays for which both decay products were measured.

The corresponding number of 'background' events is given by:
\begin{equation}
<n_b^{acc}> = <n_b> \int_{V^{acc}} dV \rho_b = \overline{\rho}_b^{acc}
V^{acc} <n_b>,
\end{equation}
where $<n_b>$ is an average number of 'background' events and
$\overline{\rho}_b^{acc}$   is an average 'background' density in
the subspace $V^{acc}$.

However, the average number of 'background' events and the 'background'
pdf can not be directly measured due to unknown contamination of
'signal' events. The 'background' pdf, $\rho_b({\bf C})$, can be safely estimated
by 'raw background' pdf, $\rho_{rb}({\bf C})$, obtained from combinations
of tracks from different events providing that $<n_s> << <n_b>$ and
that V decay products have similar $y-p_T$ distribution to that
obtained for tracks
originating from the main vertex.

The only multiplicity which is measured is 'raw signal' multiplicity i.e.
the average multiplicity of all track pairs, which are accepted as
candidates for products of V decay, $<n_{rs}> = <n_b> + <n_s>$.
If the total 'signal' multiplicity is much smaller than total 
'background' multiplicity  one can safely write:
\begin{equation}
<n_b> \approx <n_{rs}>.
\end{equation}
The above approximation allows to calculate 'signal' multiplicity
in the given acceptance (V$^{acc}$):
\begin{equation}
<n^{acc}_s> \equiv <n^{acc}_{rs}> - <n^{acc}_b> \approx
<n^{acc}_{rs}> - \overline{\rho}_{rb} V^{acc} <n_{rs}>,
\end{equation}
providing that the uncertainty of 
the $<n^{acc}_s>$  due to the used approximation (Eq. 13)
is much smaller than $<n^{acc}_s>$ i.e.:
\begin{equation}
<n^{acc}_s> >> \frac {<n_s>} {<n_b>} <n^{acc}_b>
\end{equation}
or consequntly
\begin{equation}
\overline{\rho}_{rb}^{acc} << 1.
\end{equation}
Therefore in the region of cut--space in which the last inequality
is satisfied the 'signal' multiplicity can be expressed as:
\begin{equation}
<n_s> = <n_{rs}>(\overline{\rho}_{rs}^{acc}
- \overline{\rho}_{rb}^{acc}).
\end{equation}

For a given number of collisions, $N_{ev}$, the 'signal' can be
extracted from the 'background' in the subspace $V^{acc}$ when
the statistical uncertainty of the $<n_s>$ is much smaller than
$<n_s>$. Assuming that $<n_{rs}>$, $\overline{\rho}_{rs}^{acc}$ and
$\overline{\rho}_{rb}^{acc}$
are independent one gets:
\begin{equation}
\sigma(<n_s>) = [ (\sigma(<n_{rs}>)(\overline{\rho}_{rs}^{acc}
- \overline{\rho}_{rb}^{acc}))^2 + 
( <n_{rs}> \sigma(\overline{\rho}_{rs}^{acc}))^2 + 
( <n_{rs}> \sigma(\overline{\rho}_{rb}^{acc}))^2 ]^{1/2}
\end{equation}

Obviously the second term dominates in the above expression
(for central collisions $<n{rs}>$ is approximately fixed
and therefore $\sigma(<n_{rs}>)$ is relatively small even for
small number of collisions; number of background combinations
increarses like N$_{ev}^2$ whereas number of signal like
N$_{ev}$ and therefore $\sigma(\overline{\rho}_{rb}^{acc})$
is $\sqrt{N_{ev}}$ times smaller 
than $\sigma(\overline{\rho}_{rs}^{acc})$.
Neglecting the first and the last terms in the Eq. 18 we get:
\begin{equation}
\sigma(<n_s>) = <n_{rs}> \sigma(\overline{\rho}_{rs}^{acc}).
\end{equation}

Let us try to estimate $\sigma(\overline{\rho}_{rs}^{acc})$.
From definition we have:
\begin{equation}
\overline{\rho}_{rs}^{acc} = \frac {n^{acc}_{rs}} {n_{rs} V^{acc}}.
\end{equation}
Due to the fact that n$^{acc}_{rs} << n_{rs}$ we can assume
Poissonian flactuations of n$^{acc}_{rs}$ and therefore we get:
\begin{equation}
\sigma\overline{\rho}_{rs}^{acc} = \frac {\sqrt{n^{acc}_{rs}}} {n_{rs} V^{acc}}.
\end{equation}
Substituting n$^{acc}_{rs}$ = n$_{rs}$ $\overline{\rho}_{rs}^{acc}$  V$^{acc}$
one gets:
\begin{equation}
\sigma\overline{\rho}_{rs}^{acc} = \sqrt{ \frac 
{\overline{\rho}_{rs}^{acc}} {V^{acc}}} 
\frac {1} {\sqrt{N_{ev} <n_{rs}>}}.
\end{equation}

Therefore the minimum number of events neccesary for signal--background
separation is given by:
\begin{equation}
N_{ev} >> \frac {<n_{rs}>} {<n_{s}>^2}
\frac {\overline{\rho}_{rs}^{acc}} {V^{acc}}
\end{equation} 
\begin{center}
{\bf Acknowledgments}
\end{center}
 We are grateful to  E. Skrzypczak
 for discussions and critical reading of the manuscript. 
 The support by polish State Committee for Scientific Research under Grants
 2 0423 91 01
 and 2 0436 91 01  is also acknowledged.

\end{document}